\documentstyle[aps,prb,epsf, multicol]{revtex}
\tighten
\begin{document}
\draft

\title{
  Frequency-dependent shot noise in long disordered SNS contacts 
}
\author{         K.\ E.\ Nagaev
}
\address{
  Institute of Radioengineering and Electronics,
  Russian Academy of Sciences, Mokhovaya ulica 11, 103907 Moscow,
  Russia\\}
\date\today
\maketitle
\bigskip
\begin{abstract}

The shot noise in long diffusive SNS contacts is calculated using the
semiclassical approach. At low frequencies and for purely elastic
scattering, the voltage dependence of the noise is of the form $S_I =
(4\Delta + 2eV)/3R$. The electron-electron scattering suppresses the noise
at small voltages resulting in vanishing noise yet infinite $dS_I/dV$ at $V =
0$. The distribution function of electrons consists of a series of steps,
and the frequency dependence of noise exhibits peculiarities at $\omega =
neV$, $\omega = neV - 2\Delta$, and $\omega = 2\Delta - neV$ for
integer $n$.

\end{abstract}

\begin{multicols}{2}
\narrowtext

In recent years much work was devoted to shot noise in hybrid
superconductor (S) -- normal-metal (N) systems.\cite{Blanter-00} The key
role in these systems is played by Andreev reflections at the NS
boundaries, in which electrons incident from the normal metal are
reflected as holes transferring a charge $2e$ to the superconductor. Most
of theoretical results in this field were obtained under the assumption of
quantum-coherent transport. The shot noise in diffusive SN contacts was
shown to double at subgap voltages $eV < \Delta$ with respect to the noise
in normal contacts,\cite{deJong-94} which was interpreted as a doubling of
the effective electron charge. In SNS contacts, multiple Andreev
reflections from the contact edges were found to greatly increase the
low-voltage noise\cite{Averin-96,Cuevas-99,Bezuglyi-99}, which was
attributed to independent transfers of large charge quanta $2\Delta/eV$
times $2e$. Both effects were observed
experimentally.\cite{Jehl-99,Jehl-00,Hoss-00} The frequency dependence of
shot noise in SN contacts was shown to exhibit peculiarities at the
Josephson frequency $\omega = 2eV$ at subgap voltages and additional
singularities at $\omega = eV \pm \Delta$ at $eV >
\Delta$.\cite{Lesovik-99} The peculiarity at $\omega = 2eV$ was
already experimentally observed.\cite{Kozhevnikov-00}

It was shown recently that the shot noise in diffusive SN contacts may be
also described using the semiclassical Boltzmann--Langevin
approach.\cite{Nagaev-00b} In this approach, the doubled shot noise is
explained by the effective heating of the electron gas in the
contact by the applied voltage.\cite{heating} 
Because the SN boundary
obstructs the heat flow from the contact into the superconducting
electrode, the effective heating is stronger and the noise is larger than
in normal contacts. This approach also allows one to consider the effects
of inelastic scattering on the noise. In this paper, the semiclassical
approach is applied to the noise in long diffusive SNS
contacts.\cite{Bezuglyi-00}


Consider a narrow normal-metal microbridge connecting two massive
superconducting electrodes. The elastic mean free path $l$ of electrons in
the microbridge is assumed to be much larger than $(D/\Delta)^{1/2}$,
where $\Delta$ is the superconducting gap and $D$ is the diffusion
coefficient in the normal metal. We restrict ourselves to the case of zero
temperature. The constant voltage bias is assumed to be much larger than
the Thouless energy $\varepsilon_T = D/L^2$. Under these conditions, it is
possible to neglect the penetration of the condensate into the microbridge
and consider it just as a normal metal with a nonequilibrium distribution
function of electrons.\cite{Nazarov-96} Since the SN
boundaries have a zero electrical resistivity and the whole voltage drop
takes place in the bulk of the contact, the noise in this systems is
caused by the randomness of the impurity scattering and the semiclassical
Boltzmann--Langevin approach\cite{Kogan-69} or its quantum
extension\cite{Nagaev-00a} may be applied. The Langevin equation for the
current fluctuations reads
\begin{equation}
 \delta{\bf j}
 =
 -D
 \nabla
 \delta\rho
 -
 \sigma
 \nabla
 \delta\phi
 +
 \delta{\bf j}^{ext},
\label{basic-dj}
\end{equation}
where $\sigma$ is the electric
conductivity, $\delta\rho({\bf r})$ is the charge-density fluctuation,
$\delta\phi({\bf r})$ is the local fluctuation of the electric potential, and
the correlator of extraneous currents $\delta{\bf j}^{ext}$ is given by
$$
 \langle
  \delta j^{ext}_{\alpha}({\bf r}_1)
  \delta j^{ext}_{\beta }({\bf r}_2)
 \rangle
 _{\omega}
 =
 4\sigma
 \delta_{\alpha\beta}
 \delta( {\bf r}_1 - {\bf r}_2 )
 T_N({\bf r}_1),
$$
$$
 T_N({\bf r}, \omega)
 =
 \frac{1}{2}
 \int d\varepsilon
 \left\{
   f({\bf r}, \varepsilon)
   [
    1 - f({\bf r}, \varepsilon + \omega)
   ]
 \right.
$$
\begin{equation}
 \left.
   +
   f({\bf r}, \varepsilon + \omega)
   [
    1 - f({\bf r}, \varepsilon)
   ]
 \right\}, 
\label{basic-T_N}
\end{equation} 
where $f({\bf r}, \varepsilon)$ is the average distribution function 
of electrons,
which is almost isotropic in the momentum space because of a strong
impurity scattering.  If the frequency $\omega$ is smaller than $1/RC$,
where $R$ is the resistance of the contact and $C$ is its capacity, the
noise in the contact is obtained just by averaging $T_N$ over its
volume:\cite{Nagaev-98}
\begin{equation}
 S_I
 =
 (4/R)
 \int ( dx/L )\,
 T_N(x),
\label{basic-S_I}
\end{equation}
where $x$ is the coordinate along the contact.
To calculate the noise, one has to find first the average distribution
function $f$, which obeys the standard diffusion equation 
\begin{equation}
 D\nabla^2 f
 +
 I_{in}(\varepsilon, x)
 =
 0,
\label{basic-diffusion}
\end{equation}
with inelastic collision integral $I_{in}$. What makes the system
different from a normal contact is the boundary
\begin{figure}
 \centerline{
  \epsfxsize8cm
  \epsffile{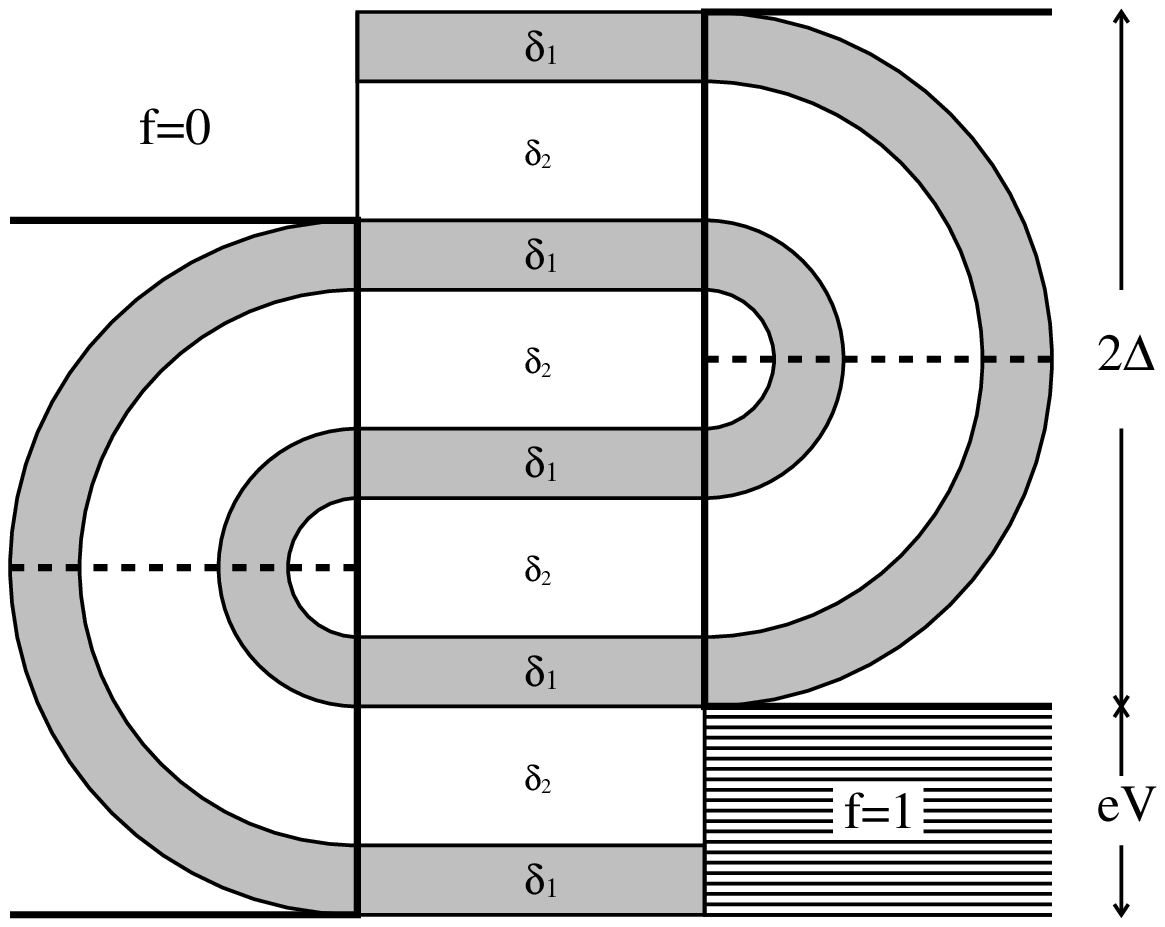}
 }
 \caption{
   The scheme of energy bands for $2\Delta/eV = 10/3$. Dashed lines 
   show the Fermi levels of the superconductors. The dashed areas show
   the path of a through passage of the quasiparticles across the 
   contact. 
 }
\end{figure}
\noindent
conditions for this
equation. Consider first a single SN boundary located at $x=0$. Assuming a
total Andreev reflection at subgap energies, one may write the boundary
conditions for the momentum-dependent distribution function at the SN
interface in the form\cite{Artemenko-77}
\begin{equation}
 f(\varepsilon, v_x)
 =
 1 - f(-\varepsilon, -v_x).
\label{basic-boundary-1}
\end{equation}
provided that the energy $\varepsilon$ is measured from the Fermi level of the
superconductor. They are conveniently
rewritten in terms of the symmetric and antisymmetric parts of $f(\varepsilon,
v_x)$ with respect to $v_x$. Expressing the antisymmetric part of $f$ in
terms of the gradient of the isotropic $f(\varepsilon, x)$, one obtains the
following boundary conditions for Eq.\ (\ref{basic-diffusion}) at the left
end of the contact:
$$
 f(\varepsilon, 0) 
 = 
 1 - f(-\varepsilon, 0),
$$
\begin{equation}
 \left.
  \nabla
  f(\varepsilon, x)
 \right|_{x = 0}
 =
 -
 \left.
  \nabla
  f(-\varepsilon, x)
 \right|_{x = 0}.
\label{basic-boundary-2}
\end{equation}
Making use of Eq.\ (\ref{basic-diffusion}), one may express the gradient
terms in (\ref{basic-boundary-2}) in terms of the distribution function at
the other end of the contact and the collision integral $I_{in}$, which
gives 
$$
  f(\varepsilon, 0)
  =
  (1/2)
  [
    1 + f(\varepsilon, L) - f(-\varepsilon, L)
  ]
$$
\begin{equation}
  +
  \frac{1}{2D}
  \int_0^L dx'
  (
    L - x'
  )
  [
    I(\varepsilon, x') - I(-\varepsilon, x')
  ].
\label{basic-boundary-3}
\end{equation}
%


Now consider the case where both electrodes are superconducting and their
Fermi levels are shifted symmetrically by $\pm eV/2$. At first assume that
the inelastic scattering is absent. Denote $f(\varepsilon, 0)$ by 
$f_L(\varepsilon)$ and
$f(\varepsilon,L)$, by $f_R(\varepsilon)$. Then one may write down the 
boundary
conditions (\ref{basic-boundary-3}) for the left and right interfaces in
the form
$$
  f_L(\varepsilon)
  =
  (1/2)
  [
    1 + f_R(\varepsilon) - f_R(-\varepsilon - eV)
  ],
$$
\begin{equation}
  f_R(\varepsilon)
  =
  (1/2)
  [
    1 + f_L(\varepsilon) - f_L(-\varepsilon + eV)
  ]
\label{elastic-boundary}
\end{equation}
%
\begin{figure}
 \centerline{
  \epsfxsize8cm
  \epsffile{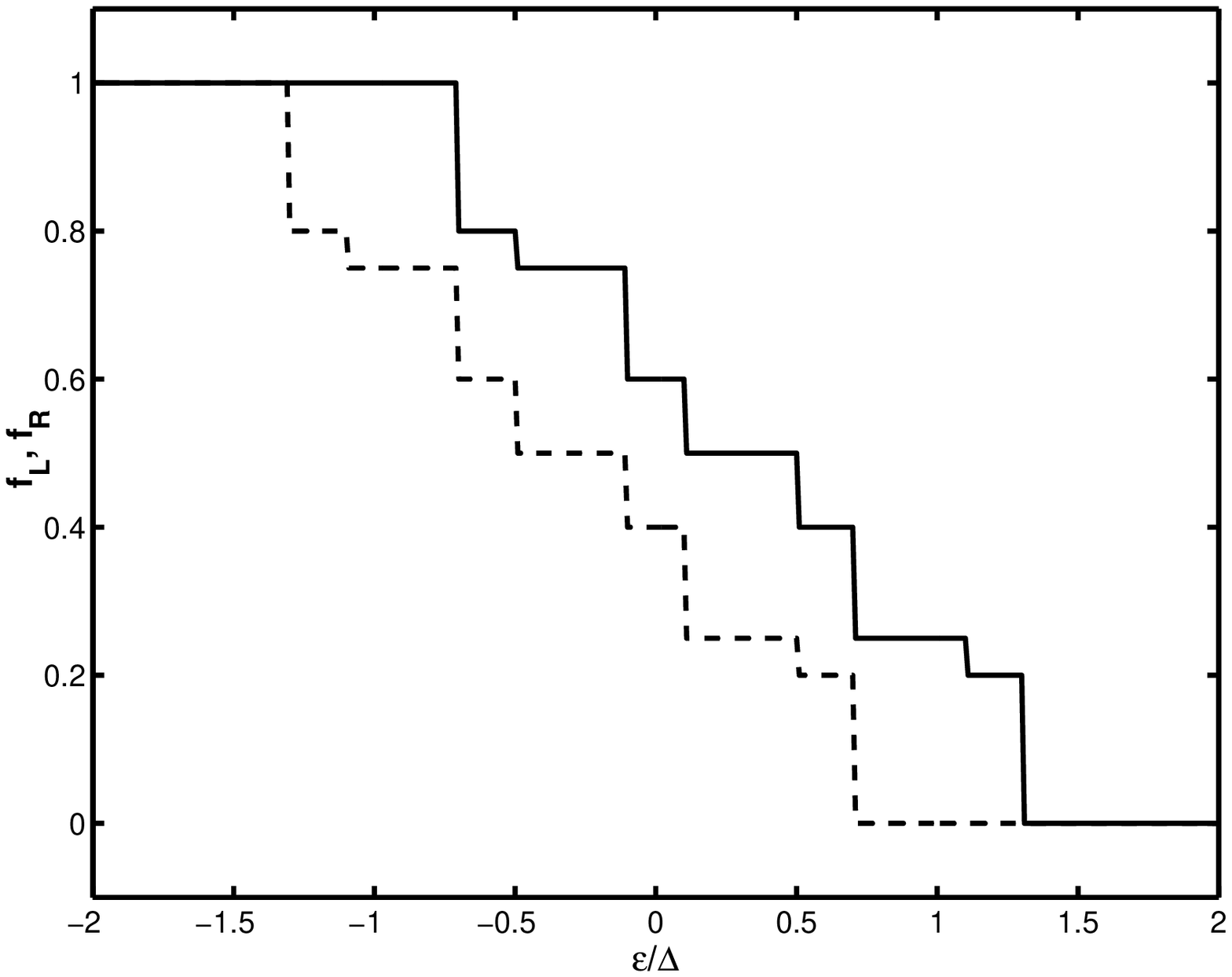}
 }
 \nopagebreak
 \caption{
   The distribution functions at the right (solid line) and left 
   (dashed line) ends of the contact for $eV/2\Delta = 3/10$.
 }
\label{FIG.2}
\end{figure}
\noindent
%

Suppose first that the ratio $2\Delta/eV = N$ is integer. In this case,
the relevant range of energies $-\Delta - eV/2 < \varepsilon < 
\Delta + eV/2$ may
be split into $N+1$ bands of width $eV$ where the distribution functions
$f_L$ and $f_R$ are constant. Denote the corresponding values by $f_{iL}$
and $f_{iR}$ in order of increasing energy, $i = 1 \ldots N+1$. It follows
from (\ref{elastic-boundary}) and the continuity of $f$ outside the subgap
region that $f_{iL}$ and $f_{iR}$ obey a set of equations
$$
   f_{iL} = ( 1 + f_{iR} - f_{N+1-i,R} )/2,
   \qquad   
   i = 1 \ldots N 
$$
$$
   f_{iR} = ( 1 + f_{iL} - f_{N+3-i,L} )/2,
   \qquad 
   i = 2 \ldots N+1,
$$
\begin{equation}
   f_{N+1,L} = 0,
   \qquad
   f_{1R} = 1.
 \label{elastic-set}
\end{equation}
It is noteworthy that at even $N$ all equations (\ref{elastic-set}) are
coupled and a quasiparticle can traverse the contact via a
series of Andreev reflections. For odd $N$ this set of
equations breaks into two independent subsets for odd and even
$i$.  In this case, the quasiparticles cannot
traverse the contact via a series of Andreev reflections and are
finally reflected back into the emitting electrode. 

Suppose now that the ratio $2\Delta/eV$ is noninteger, i.\ e.\
$2\Delta/(N+1) < eV < 2\Delta/N$. In this case, the relevant range of
energies is split into alternating bands of width $\delta_1 = 2\Delta -
NeV$ and $\delta_2 = (N+1)eV - 2\Delta$, $N+2$ bands of width $\delta_1$
and $N+1$ bands of width $\delta_2$ in all (see Fig.\ 1). The values of the
distribution functions in each set of bands $f_{i(L,R)}^{(1)}$ and
$f_{j(L,R)}^{(2)}$ obey closed sets of equations (\ref{elastic-set}) with
numbers of bands $N+2$ and $N+1$. The solutions are of the form
$$
 f_{iL}^{(1)}
 =
 1 - i/(N + 2),
 \qquad
 f_{iR}^{(1)}
 =
 1 - (i - 1)/(N + 2),
$$
\begin{equation}
 f_{iL}^{(2)}
 =
 1 - \frac{i}{N+1},
 \qquad
 f_{iR}^{(2)}
 =
 1 - \frac{i-1}{N+1}.
\label{elastic-solutions}
\end{equation}
A typical behavior of the distribution function for a noninteger ratio
$2\Delta/eV$ is shown in Fig.\ 2. It falls down from 1 to 0 through a
series of steps of unequal height. In principle, these steps could be
observed directly using tunneling spectroscopy.\cite{Pothier-97} The
distribution function at any point $x$ is calculated from the diffusion
equation as
\begin{equation}
 f(\varepsilon, x)
 =
 (1 - x/L) f_L(\varepsilon)
 +
 (x/L) f_R(\varepsilon)
\label{elastic-interpol}
\end{equation}
and integration in Eq.\ (\ref{basic-S_I}) is easily performed. At low
frequencies, one obtains 
\begin{equation}
 S_I(0)
 =
 2
 (
  2\Delta + eV
 )
 /3R.
\label{elastic-S_I}
\end{equation}
Despite the stepped distribution function, $S_I(V)$ is a perfectly smooth
curve. At large voltages, the SNS contact exhibits excess noise
$4\Delta/3eV$, which is two times larger than that of a long SN
contact.\cite{Nagaev-00b} At low voltages, it tends to the same finite
value, as it takes place for a diffusive SNS contact with a strong
barrier.\cite{Bezuglyi-99} The occurrence of the same prefactor $1/3$ in
the zero-voltage noise as in the voltage dependence of noise in
normal-metal contacts is not purely coincidental. At low voltages the
steps in the distribution function merge and it becomes a
coordinate-independent linear function of energy in the range $-\Delta <
\varepsilon < \Delta$. On the contrary, the distribution function in 
normal-metal
contacts is energy-independent linear function of coordinate in an energy
range of width $eV$. Since the integration in the expression for the
low-frequency noise is symmetric with respect to coordinate and energy, 
it gives the same prefactor in both cases.


Though the stepped structure of the distribution function does not
manifest itself in the low-frequency noise, it results in peculiarities of
its frequency dependence. The jumps of $dS_I/d\omega$ arise when the steps
of $f(\varepsilon)$ come across the steps of $f(\varepsilon - \omega)$ in 
the integrand
of Eq.\ (\ref{basic-T_N}). There are three types of peculiarities of
$S_I(\omega)$. The first one corresponds to frequencies $\omega = meV$, 
$m = 1 \ldots N+1$. The magnitude of the jump is 
\begin{equation}
 \theta_0^{(m)}
 =
 \frac{4}{3R}
 \frac{
   (N + 2)(N + 1)(2N - 3m + 3) + m^3
 }{
   ( N + 1 )^2 ( N + 2 )^2
 },
 \label{frequency-1}
\end{equation}
for $m = 1 \ldots N$ and
$
 \theta_0^{(N+1)}
 =
 (4/3R)
 ( N + 1 )^{-1}
 ( N + 2 )^{-2}.
$
The second type corresponds to the frequencies $\omega = meV + \delta_1 =
2\Delta  - (N - m)eV$, and the corresponding jumps of the derivative are
\begin{equation}
 \theta_{+}^{(m)}
 =
 \frac{2}{3R}
  \frac{
     (N + 2)(3mN - 3m^2 - 2) + (N + 2)^3 - m^3
  }{
     ( N + 1 )^2 ( N + 2 )
  }
\label{frequency-2}
\end{equation}
for $m = 1 \ldots N$ and
$
 \theta_{+}^{(N+1)}
 =
 (2/3R)
 ( N + 2 )^{-2}.
$
The third type of peculiarities takes place at $\omega = meV - \delta_1 = 
( N + m )eV - 2\Delta$ resulting in the jumps of the derivative
\begin{equation}
 \theta_{-}^{(m)}
 =
 (2/3R)
 ( N - m + 2 )^3
 /
 [
   ( N + 1 )( N + 2 )
 ]^2
\label{frequency-3}
\end{equation}
%
\begin{figure}
 \centerline{
  \epsfxsize8cm
  \epsffile{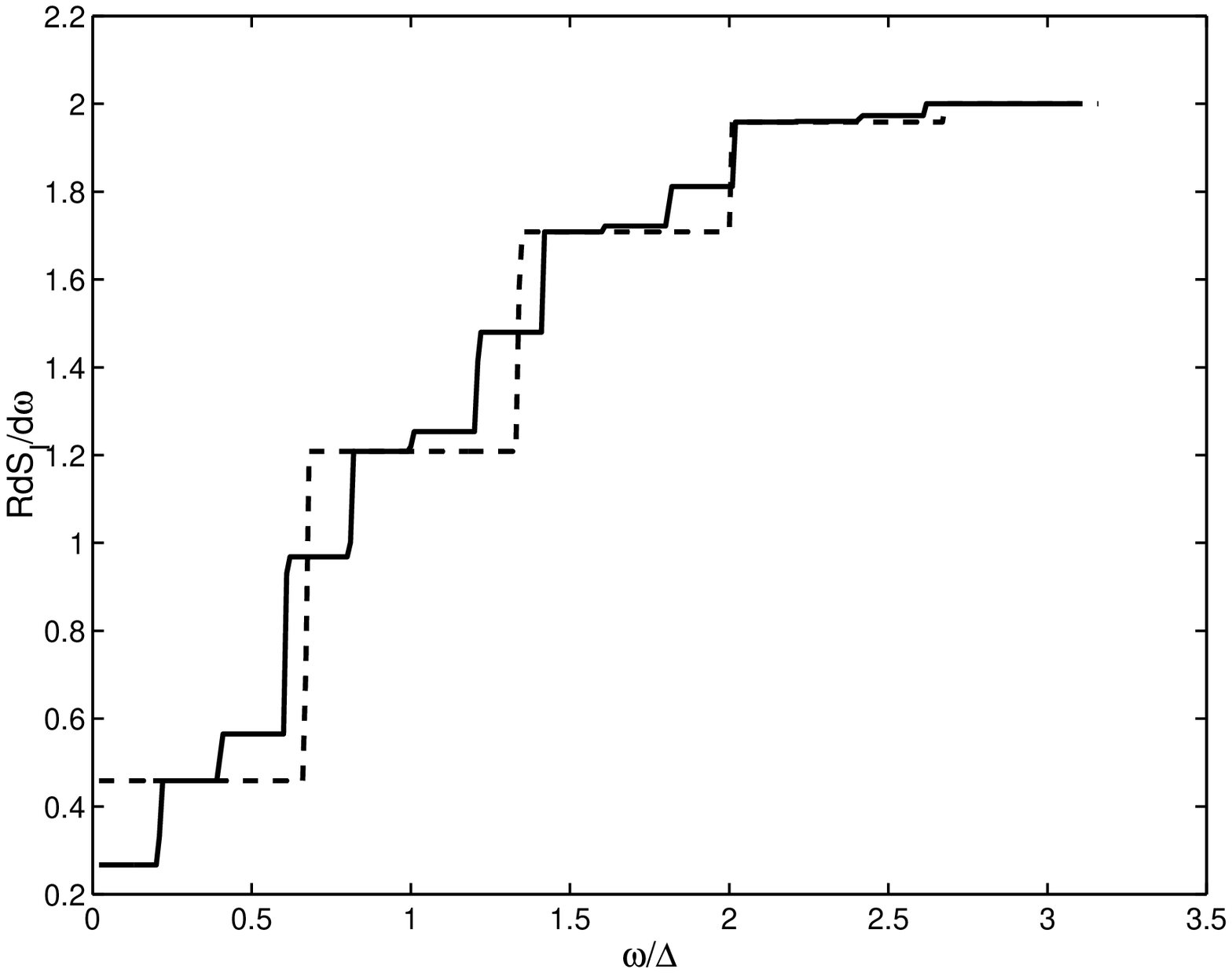}
 }
 \caption{
   Frequency dependence of $dS_I/d\omega$ for $eV/2\Delta$ $= 3/10$ 
   (solid line) and $eV/2\Delta = 1/3$ (dashed line).
 }
\label{w_steps}
\end{figure}
\noindent
for $ m = 1 \ldots N$ and
$
 \theta_{-}^{(N+1)}
 =
 (2/3R)
 [
   ( N + 1 )( N + 2 )
 ]^{-2}.
$
If $2\Delta/eV$ is integer, $\delta_1 = 0$ and the three sets of steps
merge into one at $\omega = meV$. The magnitude of the steps is given by 
the sum of $\theta_0^{(m)}$,  $\theta_{+}^{(m)}$, and  $\theta_{-}^{(m)}$ 
\begin{equation}
 \theta_{\Sigma}^{(m)}
 =
 (4/R)
 (
   N + 1 - m
 )
 /
 ( N + 1 )^2
\label{frequency-4}
\end{equation}
for $m = 1 \ldots N$ and
$
 \theta_{\Sigma}^{(N+1)}
 =
 (2/3R)
 ( N + 1 )^{-2}.
$
The frequency dependences of $dS_I/d\omega$ for an integer and noninteger
ratios $2\Delta/eV$ are shown in Fig.\ 3.


Consider now the effects of electron-electron scattering on the shot
noise. This type of scattering slightly increases it in NN and SN contacts
at $eV \ll \Delta$ because it does not decrease the energy of electron gas
yet broadens the band of partially occupied states.\cite{Nagaev-00b}
However the effects of electron-electron scattering in SNS contacts at $eV
\ll \Delta$ are essentially different from these cases since the overheated 
electron gas is trapped between two NS interfaces and quasiparticles can
escape into the electrodes only by climbing to above-gap energies through
a series of Andreev reflections or via inelastic scattering. Hence the
electron-electron scattering should decrease the shot noise because
electrons and holes outscattered from the subgap region immediately
diffuse into the electrodes and the contact is thus effectively cooled.

As the distribution function and the collision integral are only slightly
coordinate-dependent at small voltages, one obtains from Eq.\
(\ref{basic-boundary-3})
$$
  f_L(\varepsilon)
  =
  (1/2)
  [
    1 + f_R(\varepsilon) - f_R(-\varepsilon - eV)
  ]
$$
$$
  +
  (L^2/4D)
  [
    I_{in}(\varepsilon) -  I_{in}(-\varepsilon - eV)
  ],
$$
$$
  f_R(\varepsilon)
  =
  (1/2)
  [
    1 + f_L(\varepsilon) - f_L(-\varepsilon + eV)
  ]
$$
%
\begin{figure}
 \centerline{
  \epsfysize6cm
  \epsffile{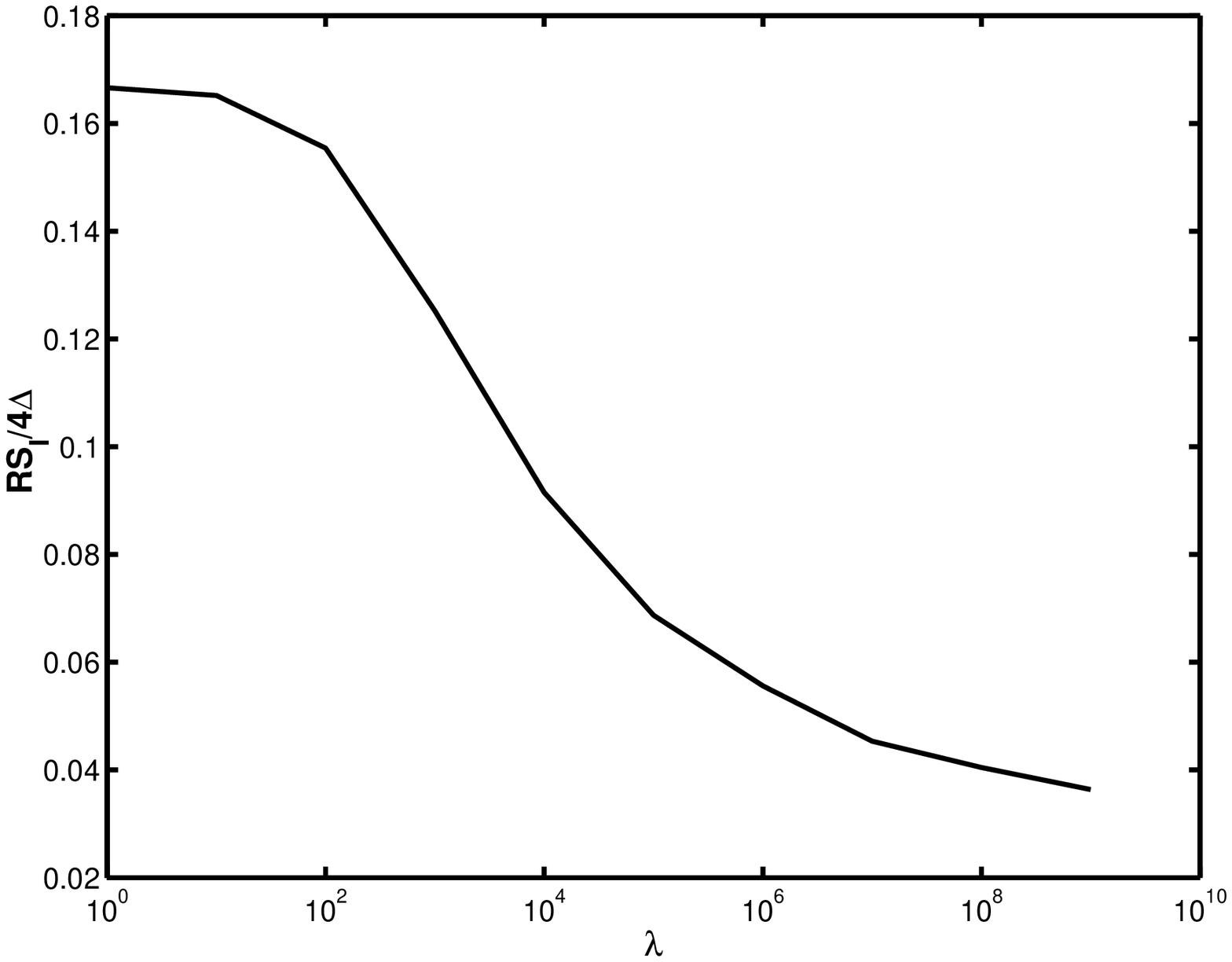}
 }
 \caption{
   Plot of the dimensionless noise vs. dimension\-less 
   parameter of electron-electron scattering at $eV \ll \Delta$.
 }
\label{S_ee}
\end{figure}
\noindent
%
\begin{equation}
  +
  (L^2/4D)
  [
    I_{in}(\varepsilon) -  I_{in}(-\varepsilon + eV)
  ].
\label{inelastic-boundary}
\end{equation}
By expanding $f_L$ and $f_R$ to the second order and $I_{in}$ to zeroth
order with respect to $eV$ and forming symmentic and antisymmetric
combinations of Eqs.\ (\ref{inelastic-boundary}) with repect to the indices
$l,R$ and $\pm\varepsilon$, one finally arrives at an energy-diffusion 
equation
for the distribution function $\overline{f}(\varepsilon) = [ f_L(\varepsilon) 
+ f_R(\varepsilon)]/2$
in the form
\begin{equation}
 (eV)^2
 d^2\overline{f}/d\varepsilon^2
 +
 (L^2/2D)
 [
   I(\varepsilon) - I(-\varepsilon)
 ]
 =
 0
\label{inelastic-diffusion}
\end{equation}
with the boundary conditions $\overline{f}(-\Delta) = 1$ and 
$\overline{f}(\Delta) = 0$.
The energy diffusion described by the right hand side of 
Eq.\ (\ref{inelastic-diffusion}) is the result of multiple Andreev
reflections in the limit $eV \ll \Delta$. In the absence of inelastic
scattering it just gives $\overline{f}(\varepsilon)= 
(1 - \varepsilon/\Delta)/2$,
which results
in $S_I = 4\Delta/3R$ in agreement with Eq.\ (\ref{elastic-S_I}).
Making use of the explicit form of the collision integral
$$
 I_{ee} (\varepsilon) 
 = 
 \alpha_{ee}           
 \epsilon_{F}^{-1} 
 \int d\varepsilon'
 \int d\omega  
 \left\{ 
    f(\varepsilon - \omega) f(\varepsilon') 
 \right.
$$
\vspace{-7mm}
$$
 \left.
   \times
   \lbrack 
      1 - f(\varepsilon) 
   \rbrack 
   \lbrack 
      1 - f(\varepsilon' - \omega) 
   \rbrack
 \right. 
$$ \begin{equation}
 \left.
    - 
    f(\varepsilon) f(\varepsilon' - \omega) 
    \lbrack 
       1 - f(\varepsilon - \omega) 
    \rbrack
    \lbrack 
       1 - f(\varepsilon') 
    \rbrack  
 \right\},
\label{I_ee}
\end{equation}
Eq.\ (\ref{inelastic-diffusion}) was solved numerically. The resulting
dimensionless low-frequency noise $RS_I/4\Delta$ is plotted in Fig.\ 4 as a
function of the dimensionless inelastic parameter $\lambda =
\alpha_{ee}\Delta^4/(e^2V^2\varepsilon_FD/L^2)$. At large $\lambda$ the noise
scales nearly as inverse logarithm of $\lambda$. This behavior may be
explained as follows. At large $\lambda$ the distribution function of
electrons trapped in the contact assumes almost equilibrium shape with an
effective temperature $T_e \ll \Delta$. On the other hand, the
quasiparticles escape from the contact in this case mostly through
electron-electron scattering rather than through Angreev reflections.
Hence the Joule heating of the contact $V^2/R$ may be equated to the flux
of energy $J$ carried by electrons and holes outscattered from the subgap
energy region by electron-electron collisions. As $J \propto
\alpha_{ee}(\Delta^3 T_e/\varepsilon_F)
\exp(-\Delta/T_e)$, one readily obtains that $S_I = 4T_e/R \propto 
R^{-1}\Delta/\ln\lambda$. This suggests that at finite $\alpha_{ee}$ the
noise tends to zero at zero voltage yet with an infinitely large slope.


In summary, we have shown that in diffusive SNS contacts much longer than
the coherence length the zero-temperature shot noise is a linear function
of voltage with excess noise two times larger than in similar SN contacts.
The finite zero-voltage noise $4\Delta/3R$ is suppressed by a finite
electron-electron scattering, but the initial slope of $S_I(V)$ remains
infinite. The stepped electron distribution function in the contact
results in three series of peculiarities of the frequency dependence of
noise at $\omega = neV$, $\omega = neV - 2\Delta$, and $\omega =
2\Delta - neV$ for integer $n$.

\end{multicols}

\begin{thebibliography}{99}
\bibitem{heating}
%
By heating, we mean a broadening of electron distribution function, which
does not necessarily imply thermalization of electrons.
\end{thebibliography}
\end{document}